\documentclass[namedreferences]{solarphysics}
\usepackage[optionalrh,solaenum]{spr-sola-addons} 
\usepackage{graphicx}                    
\usepackage{color}                       
\usepackage{url}                         

\begin{document}

\begin{article}

\begin{opening}

\title{A Standard Law for the Equatorward Drift of the Sunspot Zones}

%
\author{D.~H.~\surname{Hathaway}$^{1}$}

%
\runningauthor{Hathaway}
\runningtitle{Law of Sunspot Zone Drift}

%
  \institute{$^{1}$ NASA Marshall Space Flight Center, Huntsville, AL 35812 USA
                     email: \url{david.hathaway@nasa.gov}
             }

\begin{abstract}
The latitudinal location of the sunspot zones in each hemisphere is determined by calculating the centroid position of sunspot areas for each solar rotation from May 1874 to June 2011.
When these centroid positions are plotted and analyzed as functions of time from each sunspot cycle maximum there appears to be systematic differences in the positions and equatorward drift rates as a function of sunspot cycle amplitude.
If, instead, these centroid positions are plotted and analyzed as functions of time from each sunspot cycle minimum then most of the differences in the positions and equatorward drift rates disappear.
The differences that remain disappear entirely if curve fitting is used to determine the starting times (which vary by as much as 8 months from the times of minima).
The sunspot zone latitudes and equatorward drift measured relative to this starting time follow a standard path for all cycles with no dependence upon cycle strength or hemispheric dominance.
Although Cycle 23 was peculiar in its length and the strength of the polar fields it produced, it too shows no significant variation from this standard.
This standard law, and the lack of variation with sunspot cycle characteristics, is consistent with Dynamo Wave mechanisms but not consistent with current Flux Transport Dynamo models for the equatorward drift of the sunspot zones.
\end{abstract}

%
\keywords{Solar Cycle, Observations; Sunspots, Statistics; Sunspots, Velocity}

\end{opening}

%
 \section{Introduction} 

The equatorward drift of the sunspot zones is now a well known characteristic of the sunspot cycle.
While \inlinecite{Carrington58} noted the disappearance of low latitude spots followed by the appearance of spots confined to mid-latitudes during the transition from Cycle 9 to Cycle 10, and \inlinecite{Sporer80} calculated and plotted the equatorward drift of sunspot zones over Cycles 10 and 11, the very existence of the sunspot zones was still in question decades later \cite{Maunder03}.
The publication of the ``Butterfly Diagram'' by \inlinecite{Maunder04} laid this controversy to rest and revealed a key aspect of the sunspot cycle -- sunspots appear in zones on either side of the equator that drift toward the equator as each sunspot cycle progresses.

Cycle-to-cycle variations in the sunspot latitudes have been noted previously.
\inlinecite{Becker54} and \inlinecite{Waldmeier55} both noted that, at maximum, the sunspot zones are at higher latitudes in the larger sunspot cycles.
More recently, \inlinecite{Hathaway_etal03} found an anti-correlation between the equatorward drift rate and cycle period and suggested that this was evidence in support of flux transport dynamos \cite{NandyChoudhuri02}.
However, \inlinecite{Hathaway10} noted that all these results are largely due to the fact that larger sunspot cycles reach maximum sooner than smaller sunspot cycles and that the drift rate is faster in the earlier part of both small and large cycles.
Nonetheless, \inlinecite{Hathaway10} did find that the sunspot zones appeared at slightly higher latitudes (with slightly higher drift rates) in the larger sunspot cycles when comparisons were made relative to the time of sunspot cycle minimum.

The equatorward drift of the sunspot zones is a key characteristic of the sunspot cycle.
It must be reproduced in viable models for the Sun's magnetic dynamo and can be used to discriminate between the various models.

In the \inlinecite{Babcock61} and \inlinecite{Leighton69} dynamo models the latitudinal positions of the sunspot zones are determined by the latitudes where the differential rotation and initial magnetic fields produce magnetic fields strong enough to make sunspots.
This ``critical'' latitude moves equatorward from the position of strongest latitudinal shear as the cycle progresses.
The initial strength of the magnetic field in these models is determined by the polar field strength at cycle minimum so we might expect a delayed start for cycles starting with weak polar fields and the equatorward propagation might depend on both the differential rotation profile (which doesn't vary substantially) and the initial polar fields (which do vary substantially).

In a number of dynamo models (both kinematic and magnetohydrodynamic) the equatorward drift of the sunspot zones is produced by a ``Dynamo Wave'' (cf. \opencite{Yoshimura75}) which
propagates along iso-rotation surfaces at a rate and direction given by the product of the shear in the differential rotation and the kinetic helicity in the fluid motions.
In these models the equatorward propagation is a function of the differential rotation profile and the profile of kinetic helicity - both of which don't vary substantially.

In flux transport dynamo models (cf. \opencite{NandyChoudhuri02}) the equatorward drift is produced primarily by the equatorward return flow of a proposed deep meridional circulation.
In these models, variations in the meridional flow speed (which does vary substantially with cycle amplitude and duration in these models) should be observed as variations in the equatorward drift rate of the sunspot zones.

Here we reexamine the latitudes of the sunspot zones and find that cycle-to-cycle and hemispheric variations vanish when time is measured relative to a cycle starting time derived from fitting the monthly sunspot numbers in each cycle to a functional form for the cycle shape.

\section{The Sunspot Zones}

Sunspot group positions and areas have been measured daily since May 1874.
The Royal Observatory Greenwich carried out the earlier observations using a small network of solar observatories from May 1874 to December 1976.
The United States Air Force and National Oceanic and Atmospheric Administration continued to acquire similar observations from a somewhat larger network starting in January 1977.
We calculate the average daily sunspot area over each Carrington rotation for 50 equal area latitude bins (equi-spaced in $\sin\lambda$ where $\lambda$ is the latitude).
The sunspot zones are clearly evident in the resulting Butterfly Diagram - Figure 1.

\begin{figure} 
\centerline{\includegraphics[width=0.8\textwidth]{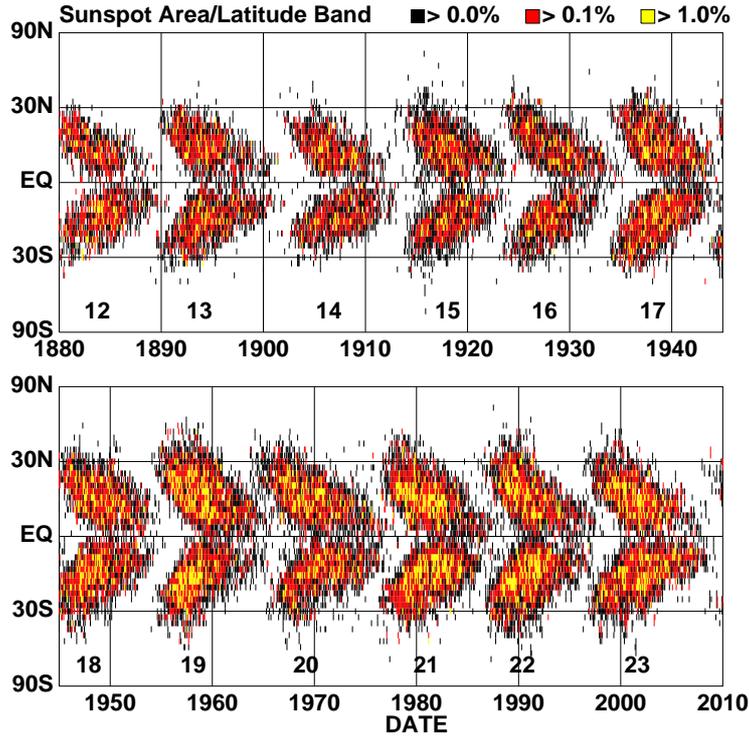}}
\caption{Sunspot areas as functions of sin(latitude) and time for each Carrington rotation from 1880 to 2010.
These data include four small cycles (Cycles 12, 13, 14, and 16), four average cycles (Cycles 15,17, 18, and 20), and four large cycles (Cycles 19, 21, 22, and 23).}
\end{figure}

We divide the data into separate sunspot cycles by attributing low-latitude groups to the earlier cycle and high-latitude groups to the later cycle when the cycles overlap at minima.
We further divide the data by hemisphere and then calculate the latitude, $\bar{\lambda}$, of the centroid of sunspot area for each hemisphere for each rotation of each sunspot cycle using

\begin{equation}
 \bar{\lambda} = \sum{A(\lambda_i) \lambda_i} / \sum{A(\lambda_i)}
\end{equation}

\noindent
where $A(\lambda_i)$ is the average daily sunspot area in the latitude bin centered on latitude $\lambda_i$ and the sums are over the 25 latitude bins for a given hemisphere and Carrington rotation.
These centroid positions then provide the sunspot zone latitudes and drift rates for each hemisphere as a function of time through each cycle.

\section{The Sunspot Zone Equatorward Drift}

Previous work on cycle-to-cycle variations in the positions and drift rates of the sunspot zones \cite{Becker54,Waldmeier55,Hathaway_etal03} made those measurements relative to the sunspot cycle maxima.
The centroid position data are plotted as functions of time from cycle maxima in Figure 2.
The data encompass 12 sunspot cycles which, fortuitously, include four cycles much smaller than average (Cycles 12, 13, 14, and 16 with smoothed sunspot cycle maxima below 90), four cycles much larger than average (Cycles 18, 19, 21, and 22 with smoothed sunspot cycle maxima above 150), and four cycles close to the average (Cycles 15, 17, 20, and 23).

Figure 2 illustrates why the earlier studies concluded that larger cycles tend to have sunspot zones at higher latitudes.
The centroid positions for the large cycles (in red) are clearly at higher latitudes than those for the medium cycles which, in turn, are at higher latitudes than those for the small cycles.
While this conclusion is technically correct, it is somewhat misleading since large cycles reach their maxima sooner than small cycles (the ``Waldmeier Effect'' \opencite{Waldmeier35} and \opencite{Hathaway10}) and the sunspot zones are always at higher latitude earlier in each cycle.

\begin{figure} 
\centerline{\includegraphics[width=0.8\textwidth]{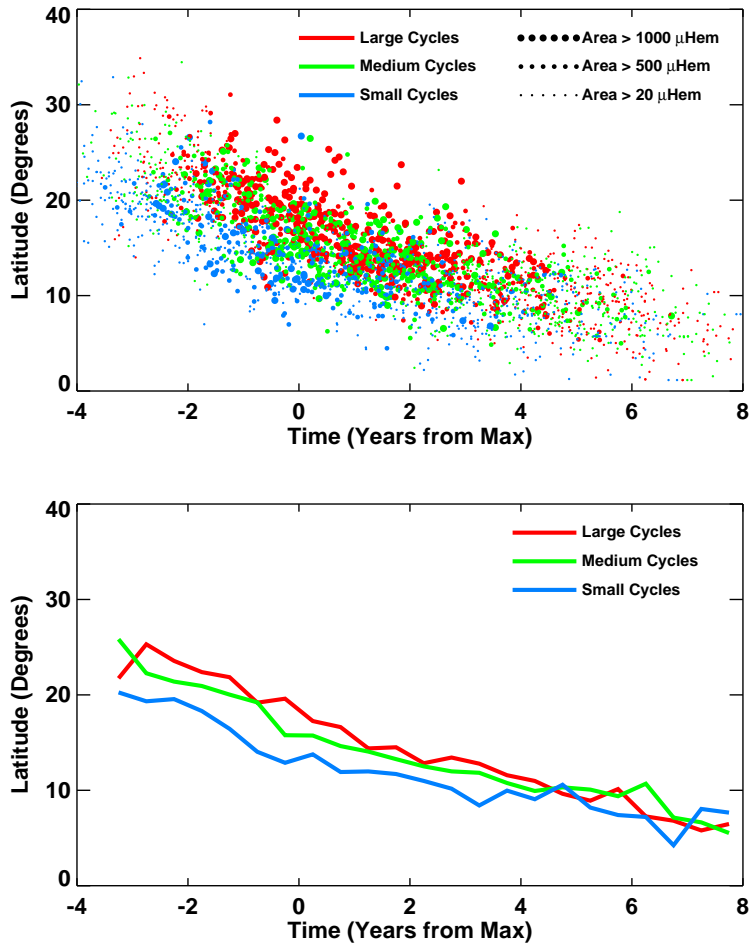}}
\caption{The centroid (area weighted) positions of the sunspot zones in each hemisphere for each solar rotation are plotted as functions of time from each sunspot cycle maximum.
The individual data points are shown in the upper panel.
The size of the symbol varies with the average daily sunspot area for each solar rotation and hemisphere.
The color of the symbol varies with the amplitude of the sunspot cycle associated with the data.
The average centroid positions of the sunspot zones for small (blue) medium (green) and large (red) cycles plotted at 6-month intervals in time from sunspot cycle maximum are shown in the lower panel.}
\end{figure}

In Figure 3 the centroid positions are plotted as functions of time from sunspot cycle minima.
Since large cycles reach maximum earlier than small cycles, the data points for the large cycles are shifted to the left (closer to minimum) relative to the medium and small cycles.
The size of the shift is different for each cycle depending on the dates of minimum and maximum.
Comparing Figures 2 and 3 shows that: 1) the latitudinal scatter is smaller in Figure 3 than in Figure 2 and; 2) the differences in the centroid positions for the different cycle amplitudes are diminished in Figure 3.
This suggests that there is a more general, cycle amplitude independent, law for the latitudes (and consequently latitudinal drift rates) of the sunspot zones.
A slight additional shift in the adopted times for sunspot cycle minima (with earlier times for small cycles) would appear to further diminish any cycle amplitude differences.

\begin{figure} 
\centerline{\includegraphics[width=0.8\textwidth]{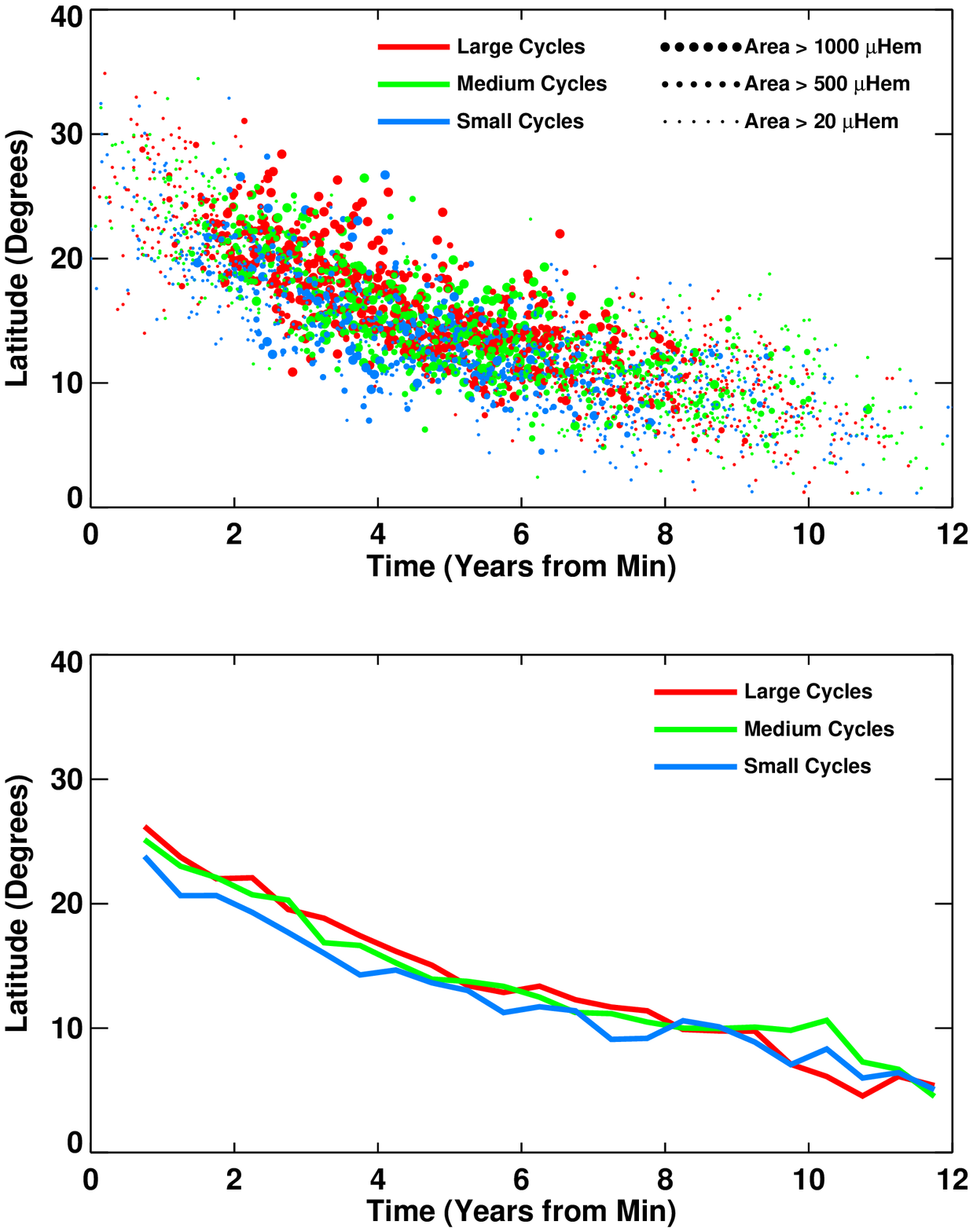}}
\caption{The centroid positions of the sunspot zones in each hemisphere for each solar rotation are plotted as functions of time from sunspot cycle minimum with the same method as in Figure 2.}
\end{figure}

Determinations of the dates of sunspot cycle minima are not well defined.
Many investigators simply take the date of minimum in some smoothed sunspot cycle index (e. g. sunspot number, sunspot area, 10.7 cm radio flux).
Unfortunately, this can give dates that are clearly not representative of the actual cycle minima.
This problem led earlier investigators to define the date of minimum as some (undefined) average of the dates of: 1) minimum in the monthly sunspot number; 2) minimum in the smoothed monthly sunspot number; 3) maximum in the number of spotless days per month; 4) predominance of new cycle sunspot groups over old cycle sunspot groups \cite{Waldmeier61,McKinnon87,HarveyWhite99}.
Even neglecting the fact the the nature of the average is not defined, it is clear from the published dates for previous cycle minima that the first criterium is never used (probably due to the wide scatter it gives) and that even the simple average of the remaining criteria doesn't give the published dates \cite{Hathaway10}.

An alternative to using this definition for the dates of sunspot cycle minima is to use curve fitting to either the initial rise of activity or to the complete sunspot cycle.
Curve fitting is less sensitive to the noise associated minimum cycle behavior (e.g. discretized data and missing data from the unseen hemisphere).
\inlinecite{Hathaway_etal94} described earlier attempts at fitting solar cycle activity levels (monthly sunspot numbers) to parameterized functions and arrived at a function of just two parameters (cycle starting time $t_0$ and cycle amplitude $R_{max}$) as the most useful function for characterizing and predicting solar cycle behavior.
This function:

\begin{equation}
F(t;t_0,R_{max}) = R_{max} \ 2 ({t - t_0 \over b})^3/\left[exp({t - t_0 \over b})^2 - 0.71\right]
\end{equation}

\noindent
is a skewed Gaussian with an initial rise that follows a cubic in time from the starting time (measured in months).
The width parameter, $b$, is a function of cycle amplitude $R_{max}$ that mimics the ``Waldmeier Effect.''
This function is

\begin{equation}
b(R_{max}) = 27.12 + 25.15(100/R_{max})^{1/4}
\end{equation}

\noindent
Fitting $F(t;t_0,R_{max})$ to the monthly averages of the daily International Sunspot Numbers using the Levenberg-Marquardt method \cite{Press_etal86} gives the amplitudes and starting times given by \inlinecite{Hathaway_etal94} and reproduced in Table 1 with the addition of results for Cycle 23.

On average the small cycles have starting times about 7 months earlier than minimum while medium cycles and large cycles have starting times about equal to minimum.
However, since minimum is determined by the behavior of both the old and the new cycles, there are substantial differences between the dates of minima and the starting times even among the medium and large cycles.
For example, Cycles 21 and 22 were both large but the minimum was 3 months earlier than the starting time in Cycle 21 and 4 months later in Cycle 22.
This is illustrated in Figure 4.

\begin{figure} 
\centerline{\includegraphics[width=0.8\textwidth]{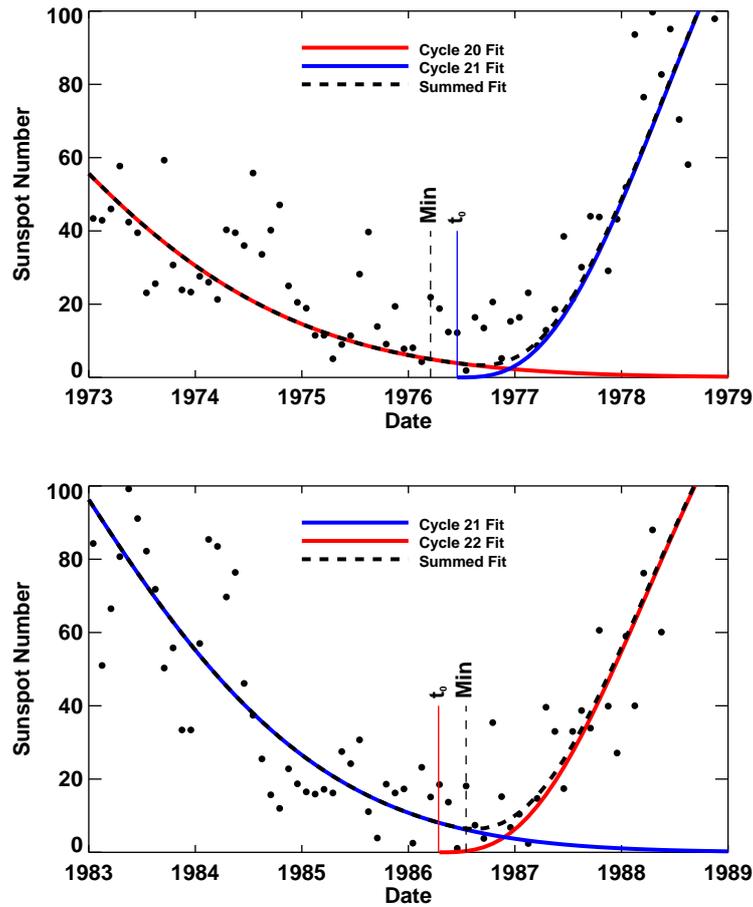}}
\caption{Monthly sunspot numbers for the Cycle 20/21 minimum (top) and Cycle 21/22 minimum (bottom).
The curves fit to each cycle are shown with the colored lines with the sum of both contributions indicated by the dashed black line.
Dates of minima and starting times are indicated to illustrate the differences.}
\end{figure}

Measuring the time through each cycle relative to these starting times (rather than minimum or maximum) removes the scatter and cycle amplitude dependence on the centroid positions as shown in Figure 5.

\begin{figure} 
\centerline{\includegraphics[width=0.8\textwidth]{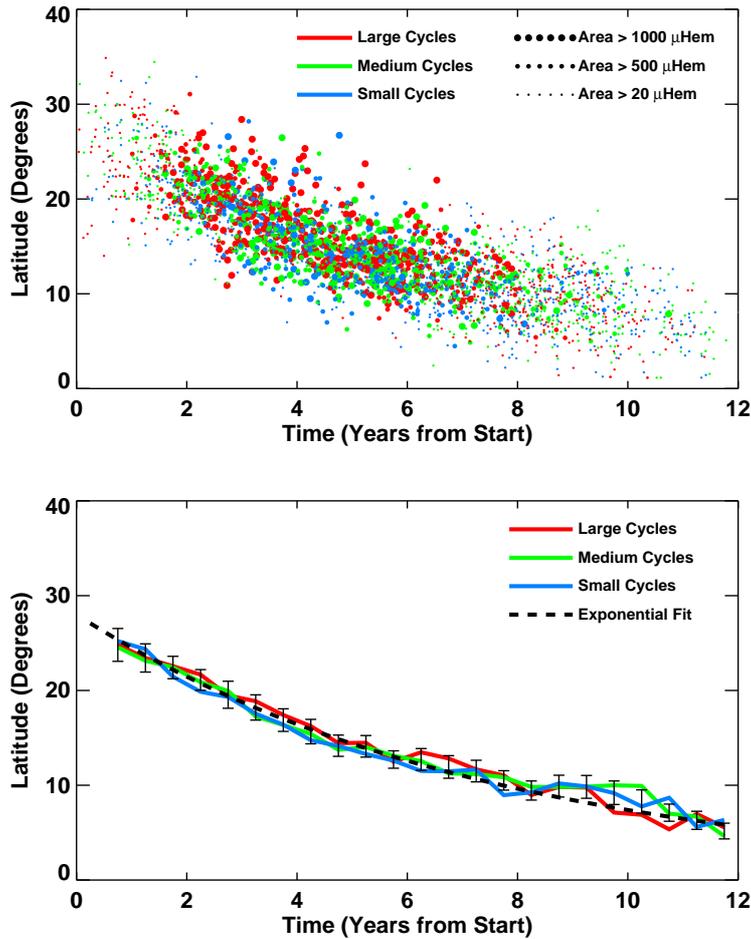}}
\caption{The centroid positions of the sunspot zones in each hemisphere for each solar rotation are plotted as functions of time from sunspot cycle start as determined by fitting a parameterized function to each cycle.
The individual data points are shown in the upper panel.
The average centroid positions of the sunspot zones for small (blue) medium (green) and large (red) cycles plotted at 6-month intervals in time from sunspot cycle start are shown in the lower panel.
The average centroid positions for all of the data are shown with $2\sigma$ error bars.
All three curves fall within the $2\sigma$ errors, criss-crossing each other along a common, standard trajectory given by the exponential fit in Equation 4 (dashed line)}
\end{figure}

%
\begin{table}
\caption{Sunspot cycle number, amplitude, minimum date, starting date, difference (starting date - minimum date in months), and dominant hemisphere.}
\begin{tabular}{cccccc}     
\hline
Cycle & Amplitude & Min & $t_0$ & $\Delta$ & Hemisphere \\
\hline
12 & 75 (small) & 1878/12 & 1878/05 & -7 & South \\
13 & 88 (small) & 1890/01 & 1889/05 & -8 & South \\
14 & 64 (small) & 1901/12 & 1901/08 & -4 & Balanced \\
15 & 105 (medium) & 1913/06 & 1913/03 & -3 & North \\
16 & 78 (small) & 1923/09 & 1923/02 & -7 & North \\
17 & 119 (medium) & 1933/10 & 1933/11 & +1 & Balanced \\
18 & 151 (large) & 1944/02 & 1944/03 & +1 & South \\
19 & 201 (large) & 1954/04 & 1954/04 & 0 & North \\
20 & 111 (medium) & 1964/10 & 1964/11 & +1 & North \\
21 & 164 (large) & 1976/03 & 1976/06 & +3 & Balanced \\
22 & 158 (large) & 1986/07 & 1986/03 & -4 & Balanced \\
23 & 121 (medium) & 1996/08 & 1996/08 & 0 & South \\
\hline
\end{tabular}
\end{table}

The lack of any substantial cycle amplitude dependence on the centroid positions when time is measured relative to the curve fitted cycle starting time suggests that the equatorward drift of the sunspot zones follows a standard path or law.
This path is well represented by an exponential function with

\begin{equation}
\bar{\lambda}(t) = 28^\circ \exp\left[-(t - t_0)/90\right]
\end{equation}

\noindent
where time, $t$, is measured in months.
This exponential fit and the data for the small, medium, and large cycles are plotted as functions of time from the cycle starting time in the lower panel of Figure 5.

Hemispheric differences in solar activity were first noted by \inlinecite{Sporer89} not long after the discovery of the sunspot cycle itself.
Much has been made of these differences and their possible connection to a variety of sunspot cycle phenomena.
\inlinecite{NortonGallagher10} recently revisited these connections and found little evidence for any of them.
Nonetheless we are compelled to examine possible differences in the sunspot zone locations and equatorward drift relative to the hemispheric asymmetries.
We keep the same starting time for each hemisphere of each cycle as determined from the curve fitting of the sunspot numbers but separate the data by the strength of the activity in the hemisphere.
Using the data shown in \inlinecite{NortonGallagher10} for the sunspot area maximum and total sunspot area for each hemisphere in each cycle we identify cycles in which the northern hemisphere dominates as Cycles 15, 16, 19, and 20, cycles in which the southern hemisphere dominates as Cycles 12, 13, 18, and 23 with Cycles 14, 17, 21, and 22 having fairly balanced hemispheric activity.
(The relevant sunspot cycle characteristics are listed in Table 1.)
This gives 8 stronger hemispheres, 8 weaker hemispheres, and 8 balanced hemispheres.
The latitude positions of the sunspot zones for the stronger hemispheres, weaker hemispheres,and balanced hemispheres are shown separately in Figure 6.
We find no significant differences in the sunspot zone latitude positions associated with hemispheric asymmetry in spite of the fact that for the unbalanced cycles the same starting time is used for both the strong and the weak hemisphere.

\begin{figure} 
\centerline{\includegraphics[width=0.8\textwidth]{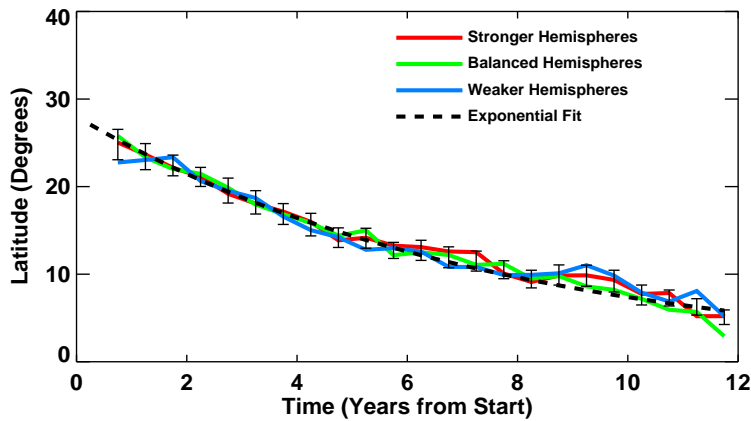}}
\caption{The average centroid positions of the sunspot zones for weaker hemispheres (blue) balanced hemispheres (green) and stronger hemispheres (red) plotted at 6-month intervals in time from sunspot cycle start.
The average centroid positions for all of the data are shown with $2\sigma$ error bars.
Here too, all three curves fall within the $2\sigma$ errors, criss-crossing each other along a common, standard trajectory given by the exponential fit in Equation 4 (dashed line).}
\end{figure}

\section{Cycle 23}

Cycle 23 had a long, low, extended minimum prior to the start of Cycle 24.
This delayed start of Cycle 24 left behind the lowest smoothed sunspot number minimum and the largest number of spotless days in nearly a century.
The polar fields during this minimum were the weakest seen in the four cycle record and the flux of galactic cosmic rays was the highest seen in the six cycle record.

One explanation for both the weak polar fields and the long cycle has been suggested by flux transport dynamos \cite{Nandy_etal11}.
This model can produce both these characteristics if the meridional flow was faster than average during the first half of Cycle 23 and slower than average during the second half.
The meridional flow measured by the motions of magnetic elements in the near surface layers (\opencite{HathawayRightmire10} and \opencite{HathawayRightmire11}) exhibited the opposite behavior - slow meridional flow at the beginning of Cycle 23 and fast meridional flow at the end.
Although the speed of the near surface meridional flow was used to estimate the speed of  the proposed deep meridional return flow in their flux transport dynamo models, \inlinecite{Nandy_etal11} suggest that the variations seen near the surface are unrelated to variations at the base of the convection zone.
However, with their model the latitudinal drift of the sunspot zones during Cycle 23 should provide a direct measure of the deep meridional flow and its variations.
  
\begin{figure} 
\centerline{\includegraphics[width=0.8\textwidth]{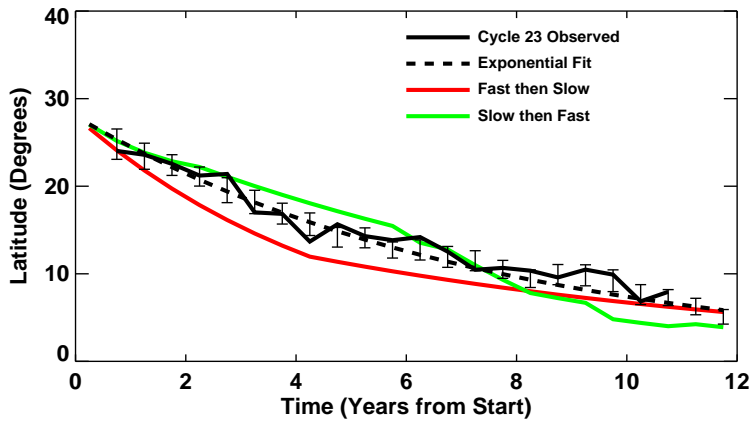}}
\caption{The average centroid positions for Cycle 23 are shown with the solid line and the exponential fit is shown with the dashed line.
The average centroid positions for all of the data are shown with $2\sigma$ error bars.
Cycle 23 data falls within the $2\sigma$ errors for the full dataset and follows the standard trajectory.
The trajectory (fast then slow) suggested by Nandy et al. (2011) is shown by the red line.
The trajectory (slow then fast) derived from the observed meridional flow variations (Hathaway \& Rightmire 2010, 2011) is shown by the green line.
}
\end{figure}

Figure 7 shows the latitudinal positions of the sunspot zones for Cycle 23 along with those for the full 12 cycle dataset (with $2\sigma$ error bars).
The latitudinal drift of the sunspot zones during Cycle 23 follows within the $2\sigma$ error range for the average of the last 12 cycles and follows the standard exponential given by Equation 4.
A drift rate that was 30\% higher than average at the start and 30\% lower than average at the end of Cycle 23 (the red line in Figure 7) as proposed by \inlinecite{Nandy_etal11} is inconsistent with the data.
A drift rate governed by the observed meridional flow variations in the near surface layers (Hathaway \& Rightmire 2010, 2011 - the green line in Figure 7) is also inconsistent with the data for Cycle 23.
This indicates that the meridional flow is not connected to the latitudinal drift of the sunspot zones.

\section{Conclusions}

We find that if time is measured relative to a cycle starting time determined by fitting the monthly sunspot numbers to a parametric curve, then the latitude positions of the sunspot zones follow a standard path with time.
We find no significant variations from this path associated with sunspot cycle amplitude or hemispheric asymmetry.

This standard behavior suggests that the equatorward drift of the sunspot zones is not produced by the Sun's meridional flow - which is observed (and theorized) to vary substantially from cycle-to-cycle.
This regularity thus questions the viability of flux transport dynamos as models of the Sun's activity cycle.
The lack of the variations in drift rate during Cycle 23 in spite of observed and theorized variations in the meridional flow also argues against these models.

The earlier kinematic dynamo models of \inlinecite{Babcock61} and \inlinecite{Leighton69} may be consistent with the regularity of the sunspot zone drift due to their dependence on the fairly constant differential rotation profile.
However, it is unclear how the variability of the initial polar fields might influence the latitudinal drift in these models.

It is clear, however, that this regularity is consistent with dynamo models in which a Dynamo Wave produces the equatorward drift of the sunspot zones.
The speed of a Dynamo Wave depends on the product of the differential rotation shear and the kinetic helicity - both of which are not observed or expected to vary substantially.

%

%
 \begin{acks}
The author would like to thank NASA for its support of this research through a grant
from the Heliophysics Causes and Consequences of the Minimum of Solar Cycle 23/24 Program to NASA Marshall Space Flight Center.
He is also indebted to Lisa Rightmire, Ron Moore, and an anonymous referee whose comments and suggestions improved both the figures and the manuscript.
Most importantly, he would like to thank the American taxpayers for supporting scientific research in general and this research in particular.
 \end{acks}

%

\end{article} 
\end{document}